# Data-driven modeling for flow reconstruction from sparse temperature measurements

**Xicheng Wang, YiMeng Chan, KinWing Wong, Dmitry Grishchenko, Pavel Kudinov**
Division of Nuclear Science and Engineering, Royal Institute of Technology, KTH
xicheng@kth.se, ymchan@kth.se, kwwo@kth.se, dmitrygr@kth.se, pkudinov@kth.se

## ABSTRACT

Measurement of the velocity field in thermal-hydraulic experiments is of great importance for phenomena interpretation and code validation. Direct measurement employing Particle Image Velocimetry (PIV) is challenging in some multiphase scenarios where the measurement system would be strongly affected by the phase interaction. In such cases, measurement can only be achieved via sparsely distributed sensors, such as Thermocouples (TCs) and pressure transducers. An example can refer to steam injection into a water pool where the rapid collapse of bubbles and significant temperature gradient make it impossible to obtain the main flow velocity at a large steam flux by PIV. This work investigates the feasibility and capability of utilization of data-driven modeling for flow reconstruction from sparse temperature data. The framework applies (i) a Proper Orthogonal Decomposition (POD) to encode variables from full space to latent space and (ii) a Fully connected Neural Network (FNN) to approximate sparse measurements to coefficients of latent space. Sensor positioning aiming to identify the optimal sensor location is also discussed. The proposed framework has been tested on a single-phase planar jet and steam condensing jets issued through a multi-hole sparger.

# 1    INTRODUCTION

Nuclear power systems require suitable tools to provide accurate state estimation for the evaluation of plant performances during normal and accident conditions. Typically, it relies on two main sources of information: (i) observations from Thermal-Hydraulic (TH) experiments and (ii) mathematic models in the form of parameterized Partial or Ordinary Differential Equations (PDEs or ODEs). The former offers valuable insights into the underlying principle of heat and mass transfer and the qualified data are utilized for the model development and validation. However, measurements are normally subject to systematic and random noise and are sparsely collected which provides limited coverage across the entire domain of interest [19]. The latter is limited by the modeling assumptions and the selection of parameters that characterize the system.

Measurement of the velocity field plays a crucial role in TH experiments for the understanding of fluid dynamics, providing valuable information for code development and validation. However, classical measurement approaches employing optical techniques such as Particle Image Velocimetry (PIV) are challenging in some applications. For instance, in the tests conducted in TALL-3D facility where the non-transparent heavy liquid metal (Lead-bismuth eutectic, LBE) served as the working fluid [1], Ultrasound Doppler Velocimetry (UDV) was originally designed to measure the velocity in the 3D test section but ultimately failed. Alternatively, hundreds of Thermocouples (TCs) were installed on the wall and internal domain of the test section, and these temperature measurements were then provided for the validation of simulation results obtained by Computational Fluid Dynamics (CFD) [2].

The challenge of optical technique also applies to the multiphase tests where the measurement system would be strongly affected by the phase interaction. For experiments conducted in PANDA facility where steam was injected through a multi-hole sparger into a subcooled water pool, PIV was implemented to measure the flow field downstream the region when steam was completely condensed [4]. The velocities at the middle and far-field regions can be partly recorded when the steam flux was relatively small at $70\ kg/m^2 s$. However, as the flux increased to $115\ kg/m^2 s$ or even higher, it became completely impractical to obtain the main flow characteristics due to the rapid collapse of bubbles and significant temperature gradient [3].

Instead of using PIV, a TC grid arranged by 42 TCs was placed in the vicinity of the sparger in PPOOLEX facility as illustrated in Figure 16. The measured temperature profiles revealed the flow characteristics to some extent and the results were then reproduced by CFD simulation using effective models in which the jets were simulated by a single-phase solver with the same amount of momentum and heat sources as created by steam condensation [28]. The velocity profiles obtained by using this effect model also achieved a good agreement with the PIV measurement in PANDA experiments at lower steam flux [6]. The simulation results indicate a similarity between the velocity and temperature profiles which can be expected since the flow after condensation is still a single-phase turbulent flow to which the Reynolds analogy can be applied. This analogy is the oldest and simplest model to estimate the turbulent Prandtl number $Pr_t$ when conducting Reynolds-averaged Navier–Stokes (RANS) simulations. It assumes a similarity between the transportation of turbulent momentum and turbulent heat transfer in a fluid.

Drawing upon the strong coupling between momentum and energy transportations, an idea has emerged to use the data-driven technique for flow velocity reconstruction by leveraging sparsely measured temperatures and data generated by dedicated mathematic models, i.e., RANS equations. This approach is expected to allow for the integration of experimental observations and mathematical models so that the fields of interest that are unavailable because of the lack of corresponding physical sensors can be indirectly determined. This concept is particularly relevant in many nuclear engineering problems in which not only velocity but also the neutron flux is strongly dependent on temperature [21].

It is worthwhile mentioning a couple of works that address velocity reconstruction through different types of observations. Velocity reconstruction can be achieved by using a framework called Physics Informed Neural Network (PINN) [7] which was designed to solve forward and inverse problems of nonlinear PDEs by the deep neural network. By applying PINN, Cai et al. [8] reconstructed the 3D flow over an espresso cup via temperatures measured by Tomographic Background Oriented Schlieren (Tomo-BOS). Di Leoni et al. [9] performed a reconstruction of Rayleigh–Bénard flow by using only temperature measurements. Additionally, the reconstruction of velocity can also be done by solving the Navier-Stokes (N-S) equations with an additional source term in the momentum equation (i.e., Boussinesq approximation) and with the solution of energy equation replaced by experimental data [10][11]. All these cases are buoyancy-driven flows in which the fluid motion is driven by a density difference due to a temperature gradient. Cammi et al [20] proposed a framework containing the Generalized Empirical Interpolation method (GEIM) and Indirect Reconstruction algorithms for a concept Circulating Fuel Reactors (CFR) to reconstruct the whole state of the system using only temperature sensors.

The goal of this paper is to study the feasibility and capability of reconstructing flow velocity of turbulent free shear flow from sparse temperature data where the heat transfer is driven by momentum transportation. We propose a data-driven framework using an encoder to perform the dimensionality reduction and a mapping function to approximate latent space coefficients through sparse sensor data. Proper Orthogonal Decomposition (POD) and autoencoder are compared for dimensionality reduction, and Fully connected Neural Networks (FNN) and linear regression are studied for the approximation. Moreover, the significance of regularization is emphasized when the input is subjected to noise.

The framework has been tested in two cases. The first case is a single-phase turbulent planar flow with training data generated by CFD. The second one is condensing jets induced by steam injection into a water pool through a multi-hole sparger in PPOOLEX facility [5]. Data obtained from scoping analysis using CFD are employed for training and temperatures measured by the TC grid are applied as input for the flow reconstruction. This work also explores the optimal sensor position that could be applied in the pre-test analysis in definition of instrumentation. It is worth noting that the framework outlined is not limited to reconstruction from temperature to velocity. Instead, it is general for the inference of inaccessible fields of interest through interdependent variables that are measurable.

The paper is organized as follows: Section 2 describes the PDE of targeted flow and the details of the applied methods; Section 3 and Section 4 present the case description and the key results of cases 1 and 2, respectively; and in Section 5 the conclusions and outlooks are drawn.

## 2 METHODOLOGY

### 2.1 Conservation equations

The conservation equations to be solved for the single-phase planar jet and jets induced by steam injection [6] are incompressible RANS equations with some simplifications as summarized in Eqs (1)~(3) [14].

$$\frac{\partial \bar{u}_i}{\partial x_i} = 0 \tag{1}$$

$$\frac{\partial \bar{u}_i}{\partial t} + \bar{u}_j \frac{\partial \bar{u}_i}{\partial x_j} = -\frac{1}{\rho}\frac{\partial \bar{p}}{\partial x_i} + (v + v_t)\frac{\partial^2 \bar{u}_i}{\partial x_j \partial x_j} \tag{2}$$

$$\frac{\partial \bar{T}}{\partial t} + \bar{u}_j \frac{\partial \bar{T}}{\partial x_j} = (\alpha + \alpha_t)\frac{\partial^2 \bar{T}}{\partial x_j \partial x_j} \tag{3}$$

where $\bar{u}_i$, $\bar{p}$ and $\bar{T}$ are time-averaged velocity, pressure and temperature, $\rho$ and $v$ the density and dynamic viscosity. $v_t$ is determined by $k/\omega$ where $k$ and $\omega$ are turbulent kinetic energy and specific dissipation rate solved by two separate equations, i.e., $k - \omega$ BSL model [14]. The terms representing body force and turbulent kinetic energy are omitted in the momentum equations (Eq. (2)).

$\alpha$ is the thermal diffusivity calculated by $= \lambda/\rho C_P$ where $\lambda$ and $C_p$ are thermal conductivity and specific heat capacity. $\alpha_t$ is the turbulent thermal diffusivity determined by $v_t/Pr_t$ via Reynolds analogy. $Pr_t$ is turbulent Prandtl number which equals to 0.85 in current turbulence model [14]. The effects of pressure work, kinetic energy, and viscous dissipation that are negligible in incompressible flows are omitted in the energy equation (Eq. (3)).

### 2.2 Data-driven framework

The data-driven framework used in this work is presented in Figure 1. From the mathematical perspective, the reconstruction problem consists in finding the relationship between sparse space ($TS[:,i] \in \mathbb{R}^s$, $i$ denotes the index of the case in steady-state simulation or snapshot in transient problem, and $s$ the number of sensors) and full velocity space ($UF[:,i] \in \mathbb{R}^n$, $n$ is the dimension of the field). Note that here we use streamwise velocity $U$ as an example. The framework can also be extended to construct full temperature space ($TF[:,i] \in \mathbb{R}^n$) and other variables as long as they are coupled with the energy equation.

Direct approximation from sparse measurement space to complete full space is challenging due to the substantial dimension difference between these two spaces. The number of TC sensors in nuclear TH experiments varies from a few to several dozen, depending on the complexity of the specific tests. For instance, in PPOOLEX facility which is a vessel with a diameter of $2.4m$ and a water level of $3m$, there are 42 TCs placed near injection orifices to capture the local phenomena (Figure 16) and an additional ~40 TCs to represent the global pool behavior [5]. However, validation using CFD code with a reasonable mesh density requires ~700,000 cells for a half domain [6].

The full fields of interest, despite complex spatial-temporal dynamics, normally exhibit low dimensional features providing the applicability of dimensionality reduction techniques. The primary objective of these techniques is to extract spatial patterns that characterize the fluid flow. If the decoding transformation is available. the reconstruction task thereby reduces to estimation of the mapping relationship from sparse space to latent space.

The data-driven framework is illustrated in Figure 1. The first step is to decode $TF$ and $UF$ by dimensionality reduction and then to estimate the mapping functions from $TS$ to $TL$ or $TS$ to $UL$. The overall workflow for the reconstruction of full temperature from sparse temperature sensors denoted as TS2TF can thereby read as $TF = d\_T \circ \mathcal{F}_1 \circ TS$. Similarly, velocity reconstruction from sparse temperature sensors denoted as TS2UF can read as $UF = d\_U \circ \mathcal{F}_2 \circ TS$. $r_1$ and $r_2$ denote the reduced dimension of temperature and velocity fields ($r_1, r_2 \ll n$). The framework is implemented in MATLAB [23] and the codes are available on GitHub at: https://github.com/xichenggege/SparseT2V.git

In this work, we compared POD and autoencoder, which are two classical techniques for linear and nonlinear encoding transformations, for dimensionality reduction and the results are presented in Section 3.2.1. For the subsequent approximation stage, non-linear regression achieved by the neural network was compared with linear regression. Furthermore, regularization was considered to enable a robust predictive capability towards input with noise. This part is reported in Section 3.2.4.

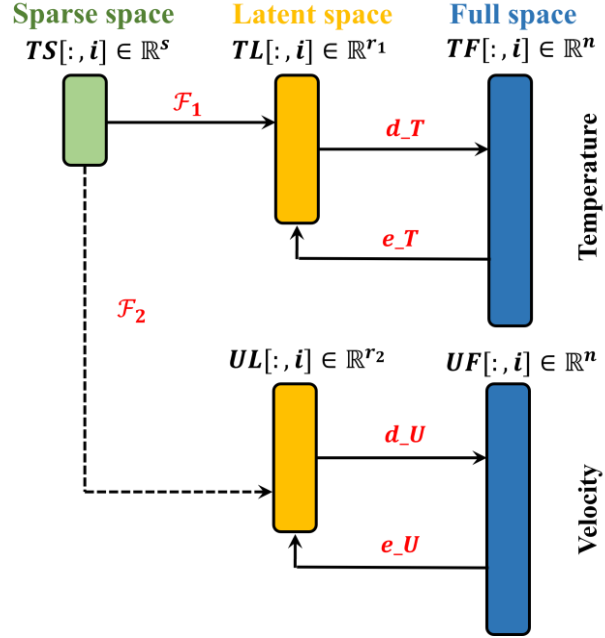

Figure 1. Representation of the reconstruction problem by the data-driven framework

## 2.3 Dimensionality reduction

The prevailing approach for dimensionality reduction is the POD which is the utilization of Principal Components Analysis (PCA) or Singular Value Decomposition (SVD) on fluid flow datasets. The resultant POD modes represent uncorrelated directions that optimally capture the variability in data. The reduction can also be achieved by several deep learning architectures such as autoencoders which contain two neural networks: an encoder to encode high dimensional data to low dimensional (latent) data and a decoder to decode latent data to recover the high dimensional results. The details of these two approaches are introduced respectively.

### 2.3.1 Proper orthogonal decomposition

The primary idea behind POD, as it was originally proposed to study turbulence in the fluid dynamic, is to decompose a fluctuated velocity field $U(\boldsymbol{r},t)$ into a collection of deterministic spatial functions $\phi_k(\boldsymbol{r})$ multiplied by time coefficients $a_k(t)$ to achieve:

$$U(\boldsymbol{r},t) = \sum_{k=1}^{\infty} a_k(t)\phi_k(\boldsymbol{r}) \qquad (4)$$

Given that time *t* is usually a pseudo-parameter [22], we replace time *t* by case index in the original definition of the POD matrix. Specifically, the training manifold of the field of interest is generated by collecting the simulation results through several steady-state (Section 3) or quasi steady-state (Section 4) simulations with varied boundary conditions and source terms. As illustrated in Figure 2, the variable of each case is reshaped into an array with $n$ dimensions, and all $m$ cases are combined to form a manifold $f$ with $n$ rows and $m$ columns.

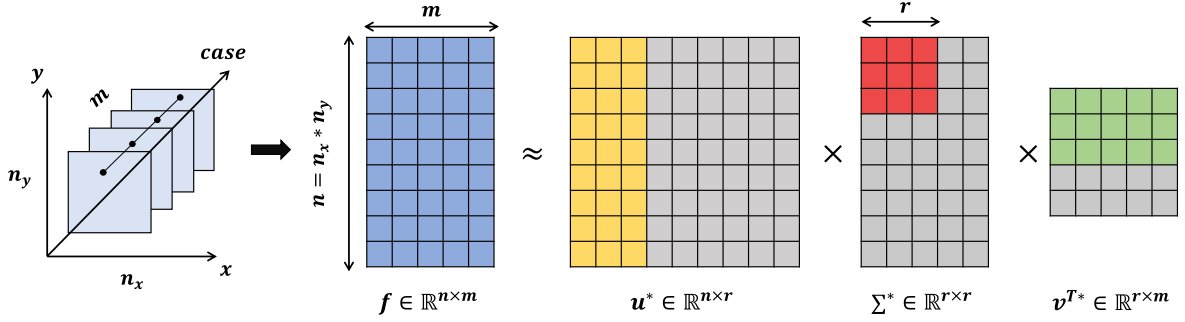

Figure 2. Schematic of POD.

The variable matrix $f$ can be decomposed by conducting SVD as:

$$f_{n\times m} = u_{n\times n}\Sigma_{n\times m}v^T_{m\times m} \approx u^*_{n\times r}\Sigma^*_{r\times r}v^{T*}_{r\times m} \tag{5}$$

where $u$ and $v$ are $n \times n$ and $m \times m$ unitary matrix, $\Sigma$ is $n \times m$ rectangular diagonal matrix. '*' denotes the disregarding the submatrices in grey (Figure 2). The matrix $f$ therefore can be approximated by the matrices with lower dimensions ($r \ll n$). It should be noted that the terms SVD and POD are usually used interchangeably in the literature. While SVD is a decomposition technique that can be used for any rectangular matrices and POD can be regarded as a decomposition formalism where SVD is one of the ways to solve its solutions. The details of SVD and its relation to POD can be found in [18].

### 2.3.2 Autoencoder

An autoencoder is a type of neural network used to copy its input to its output. Vanilla autoencoder, which is the most basic version, is applied in current work to conduct the dimensionality reduction for the variables with full dimension. It contains an encoder that compresses the high dimensional data (input) to latent representation with low dimensional and a decoder that reconstructs the input from the compressed latent representation. Both encoder and decoder are achieved by a single layer neural network with 15 hidden nodes as shown in Figure 3. The formulations for encoder and decoder are presented in Eqs. (6) and (7), respectively.

$$z = E_\Theta(X) = \sigma^1(W^1 X + b^1) \tag{6}$$

$$\tilde{X} = D_\Theta(z) = \sigma^2(W^2 \sigma^1(W^1 X + b^1) + b^2) \tag{7}$$

where $X$ and $\tilde{X}$ are input and reconstructed input, $z$ the latent space vector, $\Theta$ indicates the trainable parameters including the weights (W) and biases (b) of multiple hidden layers denoted by superscript 1 and 2. $\sigma$ is the activation function used to enable nonlinearity and the 'log-sigmoid' function, i.e., $\sigma(x) = 1/(1 + e^{-x})$, is used. The parameters $\Theta$ are trained by scaled conjugate gradient descent [23] aimed at minimizing the Mean Squared Error (MSE) between $X$ and $\tilde{X}$. The remaining training setups are set as default values as introduced in [23].

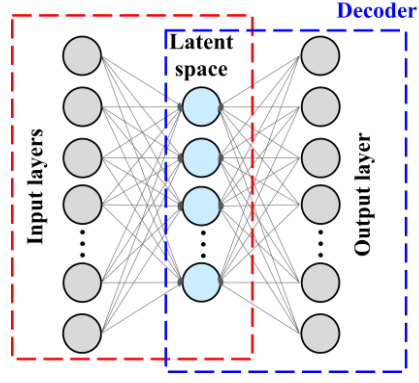

Figure 3. Schematic of Vanilla autoencoder.

## 2.4 Fully connected neural network

The neural network used for the approximation of latent coefficients from sparse measurements is a fully connected neural network (also known as feedforward neural network) with 3 hidden layers and each layer contains 10 hidden nodes. The network approximates the output $Y$ from input $X$ by conducting the calculation as presented in Eq. (8) and Figure 4.

$$\tilde{Y} = \mathcal{F}_\Theta(X) = \sigma^L(W^L \sigma^{L-1}(W^{L-1} \cdots \sigma^1(W^1 X + b^1) \cdots + b^{L-1}) + b^L) \tag{8}$$

where the definition of each parameter is the same as the one introduced in Eq. (7). The parameters Θ are trained by performing the backpropagation using the Levenberg-Marquardt algorithm with a learning rate of 1e-2 to minimize the MSE between output and reference data from the training dataset. Remaining training setups are set as default as introduced in [23]. Grid search for finding an optimal combination of these hyperparameters is recommended in future work.

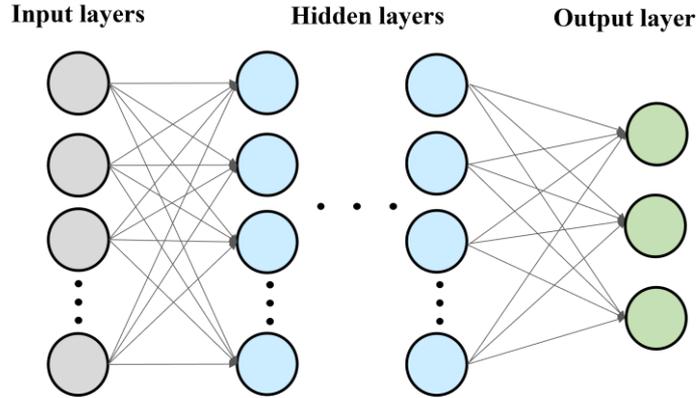

Figure 4. Illustration of full-connected neural network.

## 2.5 Linear regression

The linear regression model employed in the current work is represented by Eq. (9). In contrast to the neural network described by Eq. (8), this approach is equivalent to a single layer network with the same number of nodes as the input, while disregarding the activation function.

$$\tilde{Y} = \mathcal{F}_\Theta(X) = W^1 X + b^1 \tag{9}$$

where the definition of each parameter is the same as the one introduced in Eqs. (7) (8).

## 2.6 Regularization

Regularization is applied to improve the generalization ability of the mapping function against noisy input. For neural networks, it minimizes the performance function with a ridge penalty term (also known as L2 norm) as shown in Eq. (10). For linear regression, lasso regression and ridge regression are implemented and their performance functions are presented by Eqs. (11), (12), respectively.

$$(1-\gamma)\frac{1}{n}\sum_{i=1}^{n}(Y_i - \tilde{Y}_i)^2 + \gamma\frac{1}{n}\sum_{j=1}^{n}w_j^2 \tag{10}$$

$$\frac{1}{2n}\sum_{i=1}^{n}(Y_i - \tilde{Y}_i)^2 + \sum_{j=1}^{p}|w_j| \tag{11}$$

$$\sum_{i=1}^{n}(Y_i - \tilde{Y}_i)^2 + \sum_{j=1}^{p}w_j^2 \tag{12}$$

where $\gamma$ is a factor to weight the MSE and penalty term, $Y_i$ and $\tilde{Y}_i$ are reference and predicted values for sample $i$, $n$ is the number of samples and $p$ is the number of inputs. The parameter $w_j$ is the weight vector of length $n$ or $p$.

## 2.7 Error assessment

To assess the error between the reconstructed field and reference, Normalized Mean Square Error (NMSE) is applied. For a single case $i$, it is defined as:

$$NMSE = \frac{\left\|f(:,i) - \hat{f}(:,i)\right\|_2^2}{\|f(:,i)\|_2^2} \tag{13}$$

where $f$ and $\hat{f}$ are reference and reconstructed fields which can be temperature, U velocity, etc. By averaging the total number of cases $m$, the performance of each framework can be evaluated.

# 3 CASE1: TURBULENT PLANAR JET

## 3.1 Description of the test case

The first case is a 2D single-phase turbulent planar jet. The geometry of the case is shown in Figure 5. Data was generated using CFD code ANSYS Fluent 21.2. Simulations were conducted with an incompressible, steady-state, single-phase flow solver. The properties of water are constant. The PDEs to be solved are summarized in Eqs. (1)~(3) wherein $\partial \bar{u}_i/\partial t = 0$ and $\partial \bar{T}/\partial t = 0$. Turbulence was solved by using RANS $k - \omega$ BSL model. Energy equation was turned on while the effect of buoyancy was ignored. Temperatures at pressure outlet were fixed at 15 ℃ and non-slip condition was set to the wall.

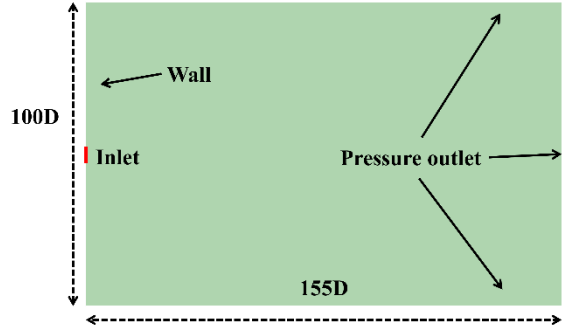

Figure 5. CFD domain of 2D planar jet.

The dataset was created through a systematic variation of the following parameters: (i) inlet velocity $U_0$, (ii) inlet temperature $T_0$, (iii) initial turbulence intensity $I_0$, (iv) initial turbulence viscosity ratio $\mu_l/\mu_t$ and (v) the magnitude of the turbulence source added to the main flow. The turbulence source is calculated by Eq. (14):

$$S_k = \begin{cases} \frac{3}{2}(v_m I_{eff})^2 \rho \frac{1}{\tau}, & v_m \geq U_0 C_{limit} \\ 0, & v_m < U_0 C_{limit} \end{cases} \quad (14)$$

where $v_m$ is cell velocity, $I_{eff}$ is introduced to relate the generation of turbulence to the cell velocity and is called "effective turbulent intensity"; $\tau$ is a characteristic time scale (assumed to be unity), $\rho$ is liquid density. $C_{limit} = 0.1$ is a filter based on the local averaged velocity that determines the region where $S_k$ should be imposed. The motivation and development of this model can be found in [6].

The variation yields a comprehensive dataset of 3740 steady-state cases that can be categorized into three groups as detailed in Table I. These parameters cover a wide range of flow conditions, particularly emphasizing the diffusion of the momentum and energy under diverse turbulent initial conditions and source terms. The framework is expected to reconstruct the flow with unknown turbulence boundaries via sparse temperature measurement. This is particularly important for the jet induced by steam injection into a water pool. The jet after complete condensation exhibits a self-similarity profile like a single-phase free shear jet but diffuses much wider [3]. The downstream velocity profiles of these two-phase phenomena can be simulated by using a single-phase solver with effective momentum, heat, and turbulence sources [6].

An example result is shown in Figure 6 in which a similarity between energy transportation and momentum transportation can be observed. This is attributed to the constant turbulent Prandtl number $Pr_t$ number that connects $v_t$ and $\alpha_t$ as presented Eqs. (2)~(3). In turbulent flows, the $v_t$ and $\alpha_t$ are typically much larger than $v$ and $\alpha$ and therefore the diffusion is dominant by the effect of turbulent.

Table I. Boundary conditions of data generation

| # Group (number of cases) | $U_0$ [m/s] | $T_0$ [°C] | $I_0$ [-] | $\mu_l/\mu_t$ [-] | $I_{eff}$[1] [-] |
|---|---|---|---|---|---|
| 1 (1300) | [0.1:0.1:10] | [15:5:80] | 0.05 | 10 | 0 |
| 2 (1120) | [1.0, 2.5, 4.0, 5.5, 6.5] | [25:15:70] | [0.05, 0.1:0.1:0.7] | [10, 500, 1000:1000:5000] | 0 |
| 3 (1320) | [1.0:0.5:6.0] | [25:15:70] | 0.05 | 10 | [0.1:0.1:3.0] |

[1] $I_{eff}$ is effective turbulence intensity to estimate the magnitude of the kinetic energy source (Eq. (14)).

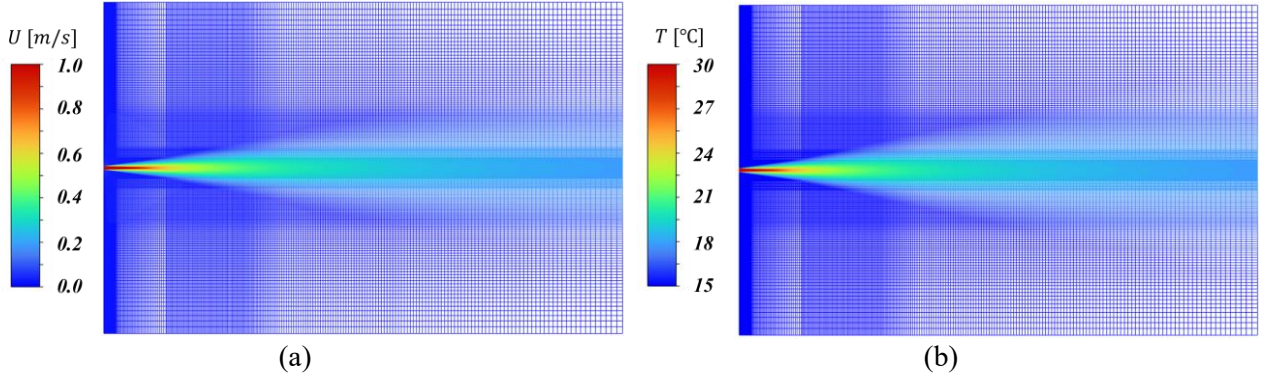

Figure 6. Contours of (a) velocity magnitude and (b) temperature obtained by CFD simulation with $U_0 = 1\ m/s$, $T_0 = 30\ °C$, $I_0 = 5\%$, $\mu_t/\mu_l = 10$, $I_{eff} = 0$.

## 3.2 Results and discussions

In this case, we focus on the reconstruction of streamwise velocity from sparsely measured temperatures. Therefore, the variables *U* and *T* are discussed mostly. To reconstruct the remaining variables, a similar manner can be applied. The training library was derived from CFD simulations through grid interpolation on a $95 \times 160$ mesh. The shape of the dataset for each variable is $15200 \times 3740$ where 15200 represents the dimensions of a single snapshot and 3740 corresponds to the total number of the steady-state cases. The dataset used for training is the fluctuating fraction of the variables and they are defined as $X = X - \bar{X}$ where $\bar{X}$ is the mean of the matrix computed over all snapshots. The dataset was divided into a training set (85%) to fit the model and a testing set (15%) to evaluate the performance of the model after training. This testing set was never exposed to the code at any stage of the process.

### 3.2.1 Dimensionality reduction

By performing POD on these matrices, the 99.9% variability (corresponding to 0.999 cumulative variances) of *T* and 99.99% variability of *U* fields can be optimally described by 11 and 13 modes as shown in Figure 7. Consequently, the snapshots with 15200 dimensions can be reduced to a lower dimension represented by $v_{r,i}^{T*}$, and reconstructed via the extracted modes by $u_{n \times r}^{*} \Sigma_{r \times r}^{*} v_{r \times m}^{T*}$ (Eq. (5)). It can also be seen from Figure 7 that the first and second modes contributes majority variability to both fields. For instance, for the velocity field, its first and second modes account for the 94.39% and 4.51% variability.

The modes extracted from the temperature field behave in a similar way to the velocity field. Specifically, both their first modes demonstrate a convection-dominant characteristic driven by the boundary conditions of velocity and temperature at the inlet. Conversely, the second modes are primarily governed by the diffusion term in their respective conservation equation. By recalling the PDEs presented in (1)~(3), the diffusivity term in the energy equation operates analogously to eddy viscosity in the momentum equations. This type of N-S equation belongs to the class of convection-diffusion equations.

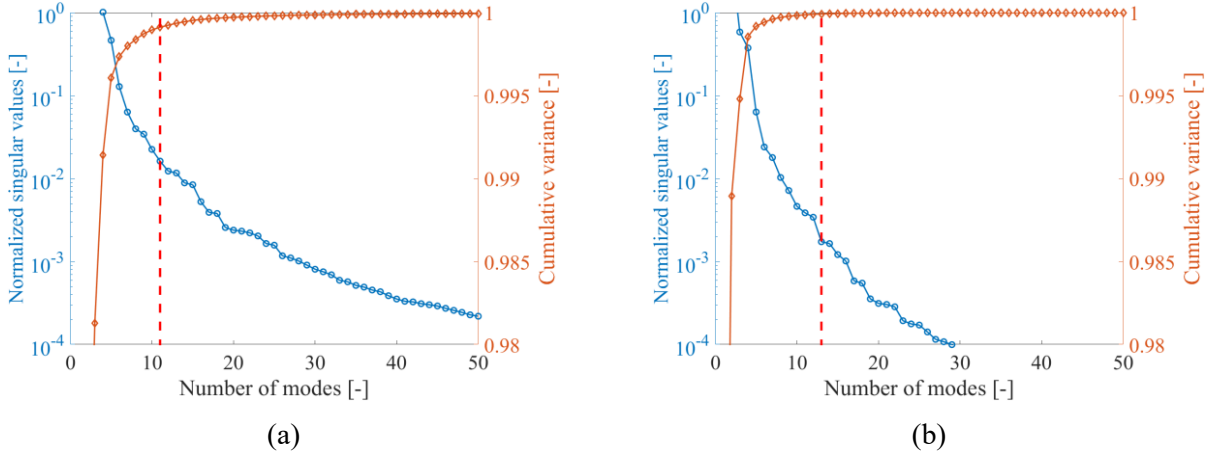

Figure 7. Cumulative variance and normalized singular values as a function of the number of modes for (a) temperature and (b) U velocity fields of planar jet case.

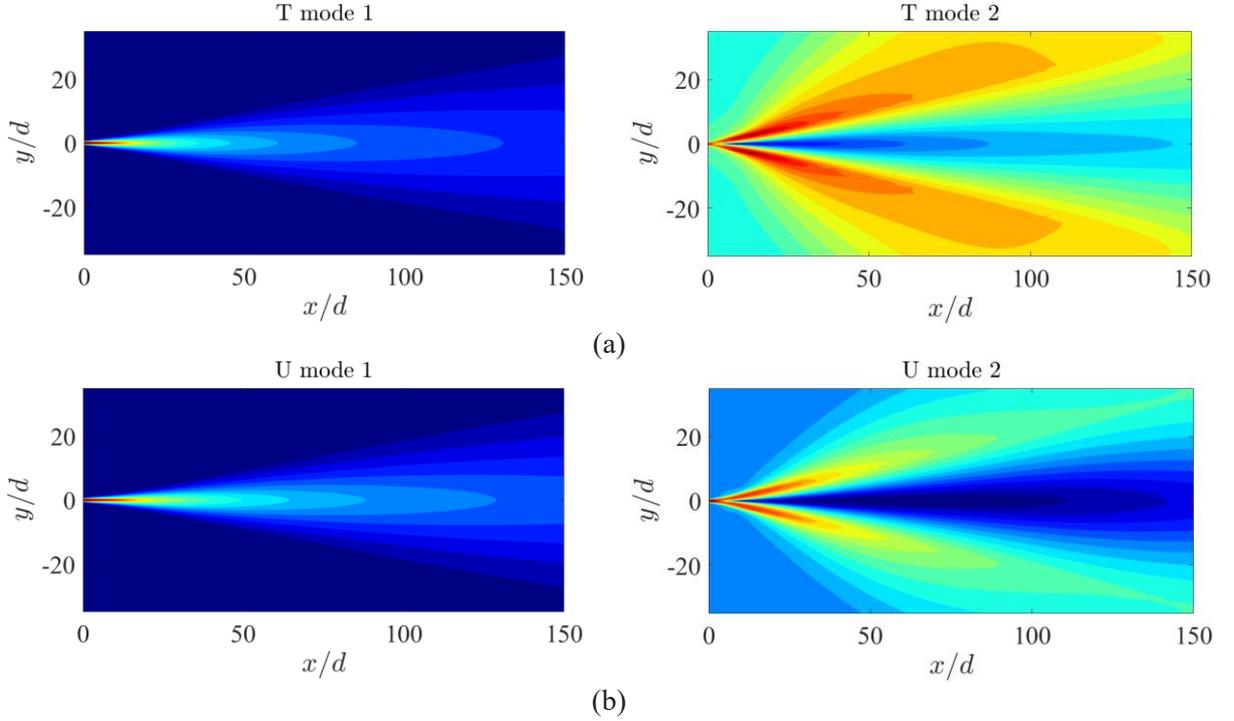

Figure 8. First and second POD modes of (a) T and (b) U fields.

The effect of noise on latent space during the decoding process is also investigated. The noise applied to the latent space of temperature (same in velocity) reads:

$$\epsilon \sim N(0, I_L[max(TL[i_r,:])]) \tag{15}$$

where the noise $\epsilon$ is assumed to follow a zero mean normal distribution and $I_L$ is the noisy intensity. The $max(TL[i_r,:])$ solves the row-wise maximum of the latent space matrix of T field.

The reconstruction errors on the test dataset of T and U fields obtained by POD and Vanilla AE against noise are compared in Figure 9. POD provides a much better and relatively robust performance in contrast with Vanilla AE. The reconstruction performances of velocity field are more sensitive to the noise added at the latent space. Optimization of network architecture and training setups, such as applying regularization, might improve the performance of autoencoder but is beyond the scope of this paper.

It should be noted that the selection of proper order reduction techniques depends on the specific problem. Notably, when dealing with the solutions of turbulent flow solved by Direct Numerical Simulation (DNS), POD would yield a few dominant modes alongside a large number of modes exhibiting similar variabilities that cannot be ignored [25]. Alternatively, nonlinear order reduction method, e.g. autoencoder using neural networks, might become a better option. Work done by Dubois et al. [12] indicates that Variational Autoencoder (VAE) in which the latent vector is sampled from the encoder-generated distribution, achieves a better reconstruction performance against noisy data compared to POD. Nevertheless, given that the reconstruction performance via POD is sufficient for the current task, it is selected for dimensionality reduction in the following work.

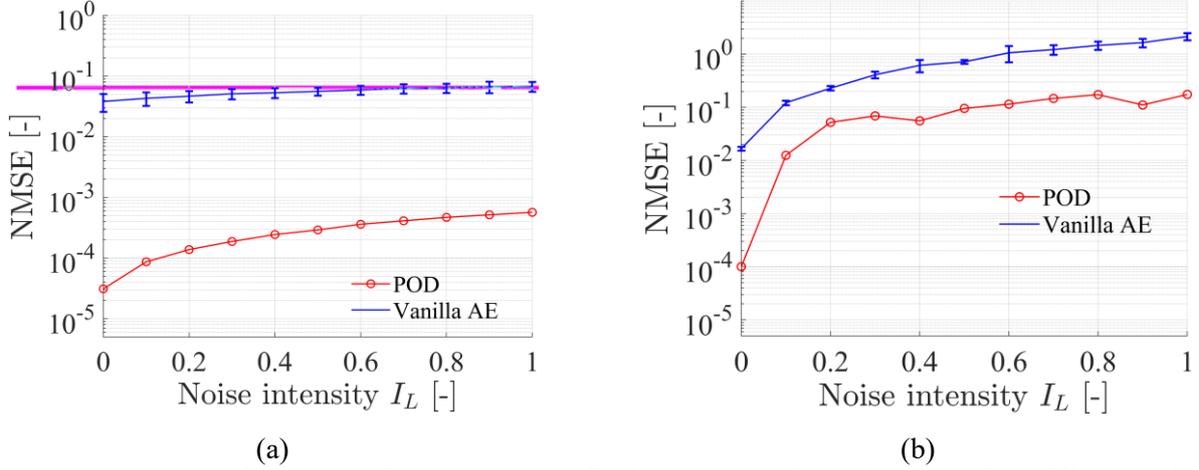

(a) (b)

Figure 9. Reconstruction errors of (a) T and (b) U fields by POD and vanilla AE against different noise intensities. The error bar of Vanilla AE represents the standard deviation (std) of 10 repeated trainings.

### 3.2.2 Sensor placement

The spatial arrangement of sensors plays a significant role in the reconstruction performance. For regressive tasks, sensors are ideally positioned to measure non-redundant signals. The sensor configuration can be determined by performing optimal sensor placement via QR decomposition as introduced in [13]. It concerns the problem of choosing a limited subset of sensors in such a way that measurement by these sensors performs nearly as well as the one measured at every point. However, it should be highlighted that these sensors are supposed to measure the same quantity as the target field. For instance, performing optimal sensor placement on the temperature field yields optimized locations only if temperature sensors are applied, i.e. TS2TF as shown in Figure 1. It becomes no longer optimized if these sensor recordings are used to infer velocity (TS2UF).

In this work, we investigated three arrangements of temperature sensors as illustrated in Figure 10 and the results are summarized in Table II. The first arrangement denoted as 'Optimal T' uses sensors only derived through optimal sensor placement on the T field. The second one 'Optimal T+U' combines sensors determined by optimal placement on both T and U fields in which the locations of optimized velocity sensors are replaced by temperature sensors. The third one 'TC grid' mimics the arrangement as the one used in PPOOLEX facility.

### 3.2.3 Reconstruction capability

The averaged normalized mean square error (NMSE, by Eq. (13)) of temperature and velocity reconstructions by 3 different sensor arrangements are compared in Table II. FNN was used to approximate sparse data to latent coefficients and each case was repeatedly trained 10 times to get sufficient statistics. Within every training, the NMSE of each test case was calculated and then averaged over across all test cases.

The temperature reconstruction from sparse data (TS2TF) achieves similar performance as to the dimensionality reduction by POD (TF2TL) on noise-free latent coefficients (Figure 9a), indicating the good prediction capability of the neural network for approximation of the latent space coefficients via temperature data (TS2TL). The cases with sensors arranged by optimal sensor placement, even with a reduced number, can still provide comparable performance to the case with TC grid-like arrangement.

However, the performance of velocity reconstruction (TS2UF) is much worse than temperature reconstruction in which their NMSE is three orders of magnitude higher than the NMSE of its dimensionality reduction (1e-1 vs 1e-4 as shown in Figure 9b). The primary reason might be the surjective but non-injective mapping between the sparse temperature space and the full velocity space. In other words, a unique velocity solution can yield different temperature profiles with varied temperatures at the inlet boundary. Using only temperature information is insufficient to infer a complete velocity field in this type of problem where the transportation of the energy is driven by the momentum.

To investigate the effect of sparse velocity data on the velocity reconstruction, the performance obtained from the cases using only temperature sensors is compared to the cases using temperature and velocity sensors as presented in Table II. While 'U orifice' and 'U downstream' refer to the velocity at the orifice and velocities obtained from the corresponding coordinates of the five furthest temperature sensors away from the orifice (Figure 10b). Involving velocity information significantly reduces the error in velocity reconstruction while using the velocity at the orifice provides better performance than using the downstream velocities. This can be attributed that the former one is the boundary condition which determine the velocity profiles at the inlet. It can be expected that the reconstruction performance can also be improved by introducing information similar to the inlet velocity, e.g. mass flow rate, as long as it can be used to identify different flow conditions.

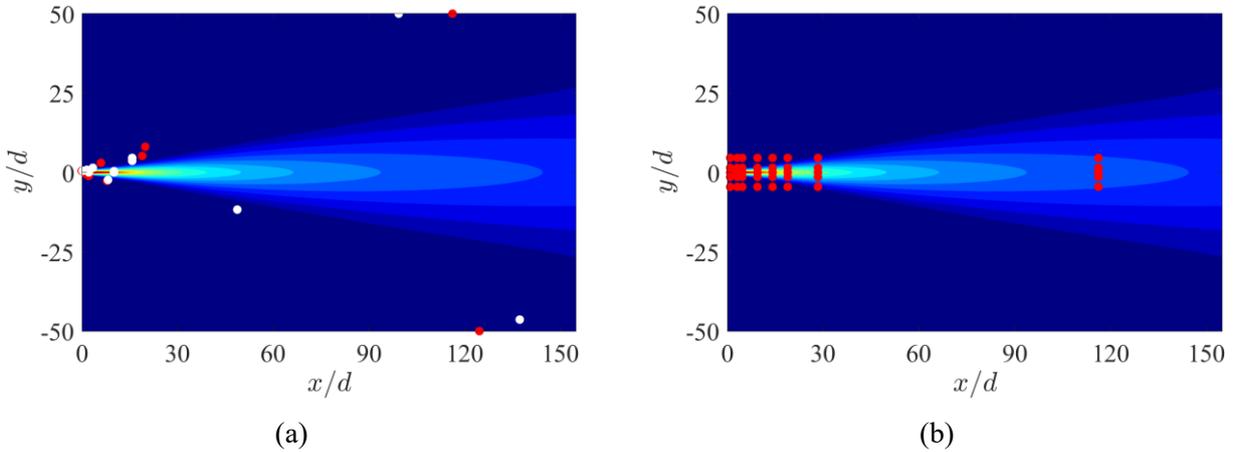

(a)          (b)

Figure 10. Temperature sensors determined by (a) optimal sensor placement on T field (red) and U field (white), and (b) using a similar arrangement as in PPOOLEX experiments (Figure 16). Visualized are the temperature profiles.

Table II. NMSE of temperature and velocity reconstructions by different sensor arrangements. The standard deviation was derived from 10 repeated trainings.

| Reconstruction type | Case | Mean of NMSE | std of NMSE |
|---|---|---|---|
| TS2TF | Optimal T | 7.6713e-5 | 1.4841e-5 |
|  | Optimal U + T | 5.1724e-5 | 1.8844e-5 |
|  | TC grid | 4.1189e-5 | 7.6163e-6 |
| TS2UF | Optimal T | 1.5806e-1 | 7.6390e-2 |
|  | Optimal U + T | 1.9940e-1 | 7.2148e-2 |
|  | TC grid | 1.6925e-1 | 1.0322e-1 |
|  | TC grid + U orifice | 5.1801e-4 | 1.4223e-4 |
|  | TC grid + U downstream | 1.2559e-3 | 6.0376e-4 |

The latent coefficients of the velocity field corresponding to primary and secondary modes are compared in Figure 11 with the values predicted by FNN through different sensor configurations and values obtained through POD. The distribution of these coefficients can serve as an indicator for the variation in boundary conditions of the dataset (Table I). The boundary conditions categorized in group 1 (corresponding to the index of 1 to 198) were varied in inlet velocity and temperature, which govern the convection of momentum and energy that can be explained by mode 1 (Figure 8). The boundaries in group 3 (index 365 to 561) differ primarily in terms of an additional turbulence source which affects the diffusion of momentum and energy that can be described by mode 2. The coefficients that cannot be correctly predicted by only using measurements from temperature sensors are well captured when velocity information is involved. This is particularly obvious for cases categorized in groups 2 and 3 (Table I) where the diffusion of the jet is governed by the initial turbulence at the inlet, or the additional turbulent source term added in the domain.

Figure 12 displays the comparison of the reconstructed velocity field through sparsely measured temperature and velocity (TC grid + U orifice) with its reference. These cases represent the three different categories of the boundary conditions in the training library (Table I). The proposed framework not only provides good predictions for the centerline velocity but also captures the downstream diffusion caused by the additional turbulence.

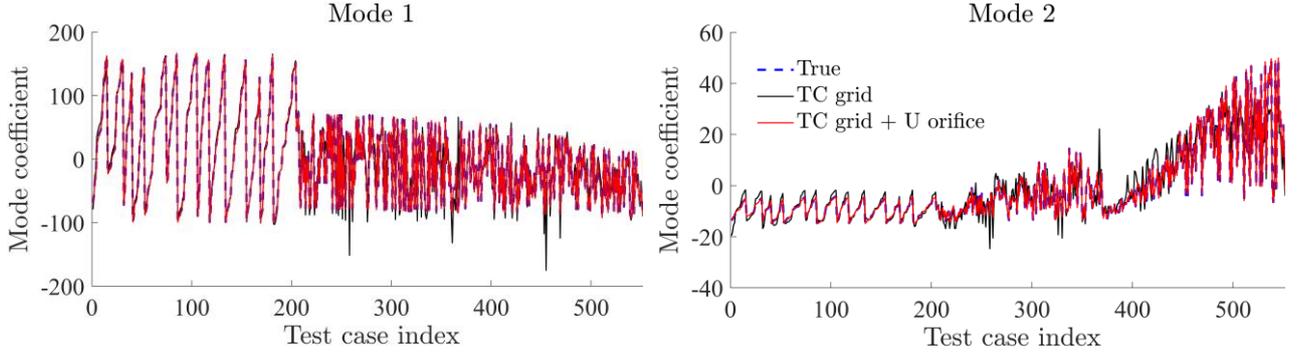

Figure 11. Comparison of coefficients of velocity latent space decoded by POD (True) and predicted with different sensor configurations.

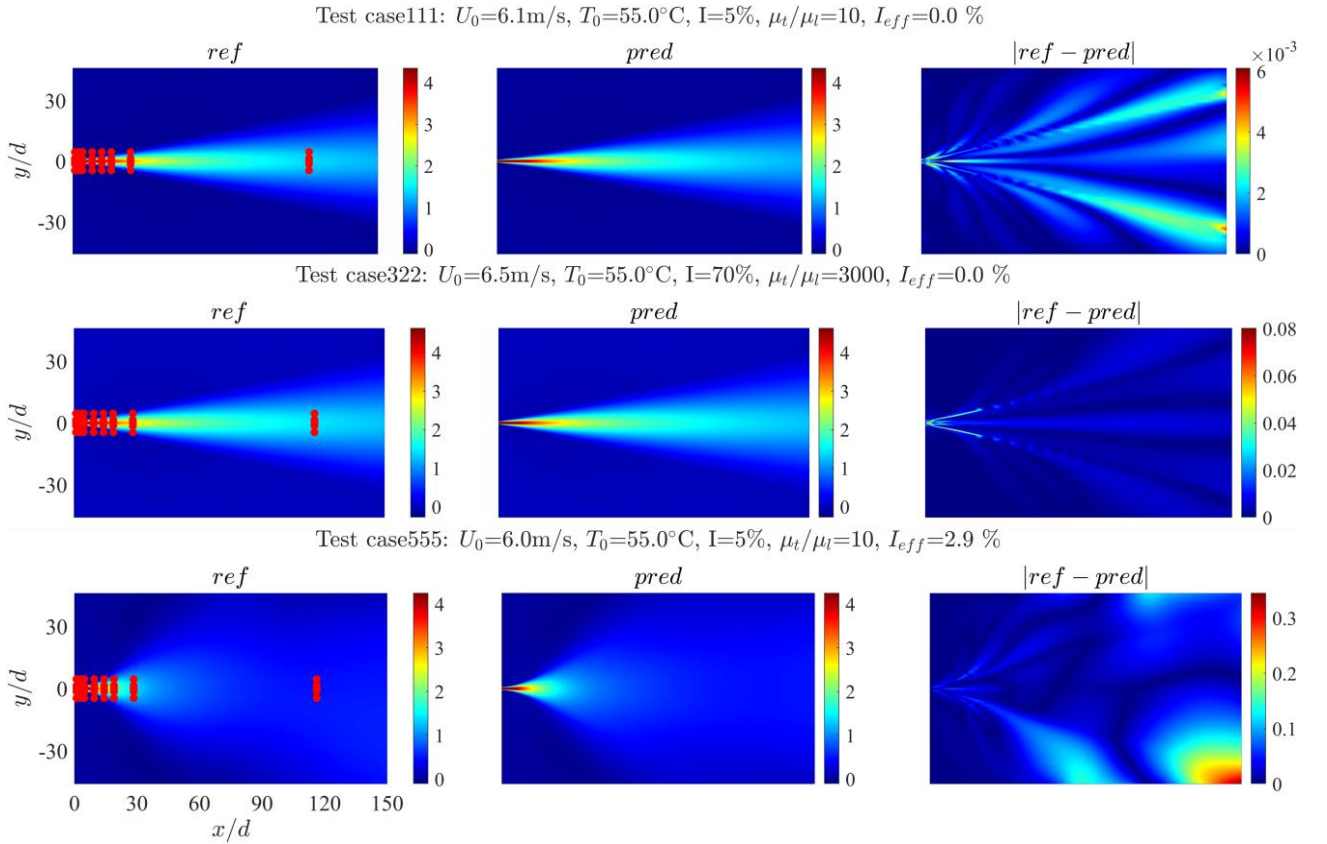

Figure 12. Examples of reconstructed streamwise velocity by sparse temperature and velocity data (TC grid + U orifice) on testing dataset.

### 3.2.4 Reconstruction against noise

It is anticipated that the reconstruction performance would be deteriorated in the presence of noise. In this case, noise may refer to the measurement uncertainty of the thermocouple. For instance, Arora et al. [26] reported a normal distribution of uncertainty with $2\sigma = \pm 2.30K$ for their K-Type thermocouple probes. The measurement noise normally consists of a ~0.1K inherent fluctuation caused by the oscillation of the electrical signal and a fixed offset varying from 0.1K to 2K during the test. This offset can be reduced by performing the calibration test. Additionally, noise can also be introduced by the discrepancy in sensor location between the experiment and data-driven input.

To assess the robustness of the proposed framework, we polluted the input of sparse temperature by introducing random noise. The noise was modeled as a random variable following a normal distribution with $N(0, \sigma^2)$. We first investigated the effects of mapping algorithms (linear or non-linear regression) and regularization on the reconstruction performance of both temperature and velocity fields. The results are compared in Figure 13. These mapping algorithms are used to approximate the corresponding latent coefficients from their sparse input. The temperature sensors were arranged as the 'TC grid' (Figure 10b) while noise-free orifice velocity is additionally involved in the velocity reconstruction.

The best prediction capability was obtained when a non-regularized neural network (i.e. $\gamma = 0.0$, see Eq. (10) for details) was applied to the noise-free input. Linear regression without regularization yields similar performance for temperature reconstruction but worse accuracy for velocity reconstruction. The prediction accuracy for both linear and non-linear approaches reduces significantly when the noise is introduced. Adding regularization terms enhance the robustness of both methods wherein the linear regression with ridge term results best performance against inputs with larger noise and the neural network with $\gamma = 0.5$ shows slightly better performance for less noisy input (i.e. $\sigma = 0.2$).

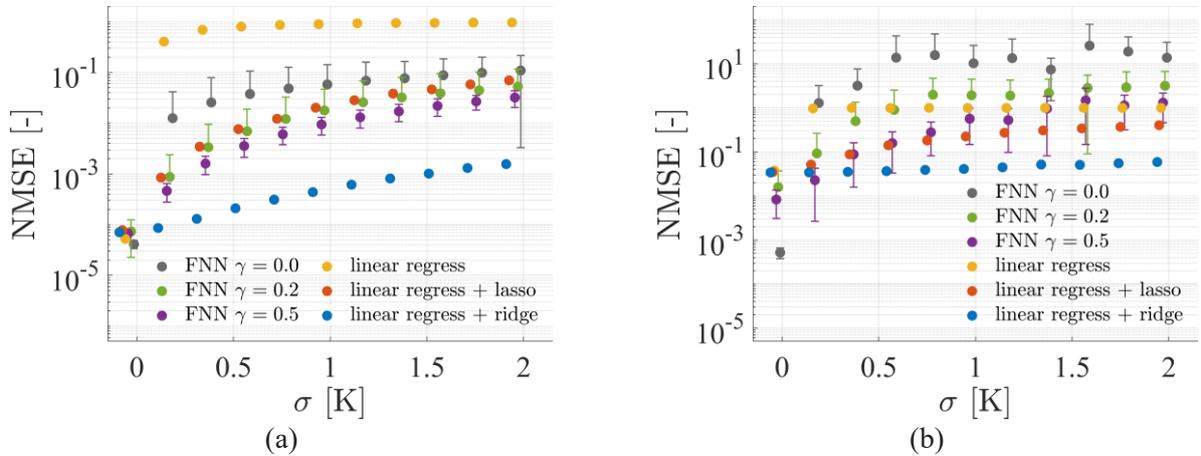

Figure 13. Performance of reconstruction of (a) temperature and (b) velocity against noisy input. Cases compared with different mapping approaches with or without regularization for TS2TL and TS2UL. The error bar represents the standard deviation (std) of 10 repeated trainings.

The reconstruction performances evaluated by using different sensor arrangements are compared in Figure 15. The neural network with $\gamma = 0.5$ is used to map the sparse inputs into the latent space coefficients. Sensors arranged as TC grid and the ones determined by optimal sensor placement (Figure 10) are compared with the sensors randomly positioned. The random sensors were obtained by random sampling within the interior domain (i.e., $|y/d| < 35 \cap x/d < 120$). The sensor number was identical to the case of 'optimal T'. This process was repeated 5 times, and their corresponding positions were displayed with distinct colors as shown in Figure 14. Note that for the velocity reconstruction, the noise-free velocity at the orifice was involved in the input for each sensor arrangement.

For noise-free input, all sensor configurations yield similar performance of temperature reconstruction while random sensors yield worse accuracy in terms of the velocity reconstruction. Sensors arranged by optimal placement demonstrate superior stability against noise in contrast with the results obtained by TC grid or random sensors. Additional, case comparison of velocity reconstruction between 'Optimal T' and 'Optimal T+U' (Figure 15b) indicates that the increasing of sensor number might deteriorate the model stability since extra sensors would also introduce additional noise.

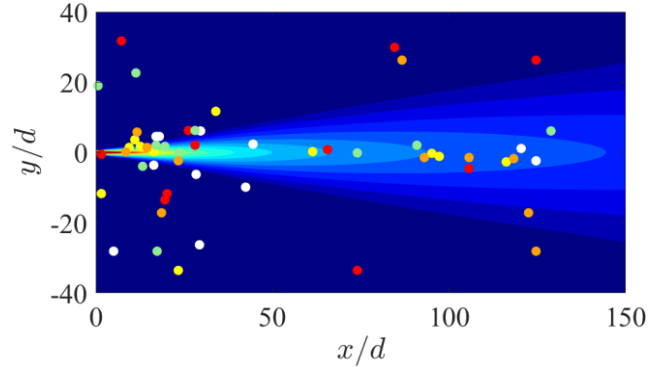

Figure 14. Temperature sensors determined by random sensor placement. Visualized are the temperature profiles.

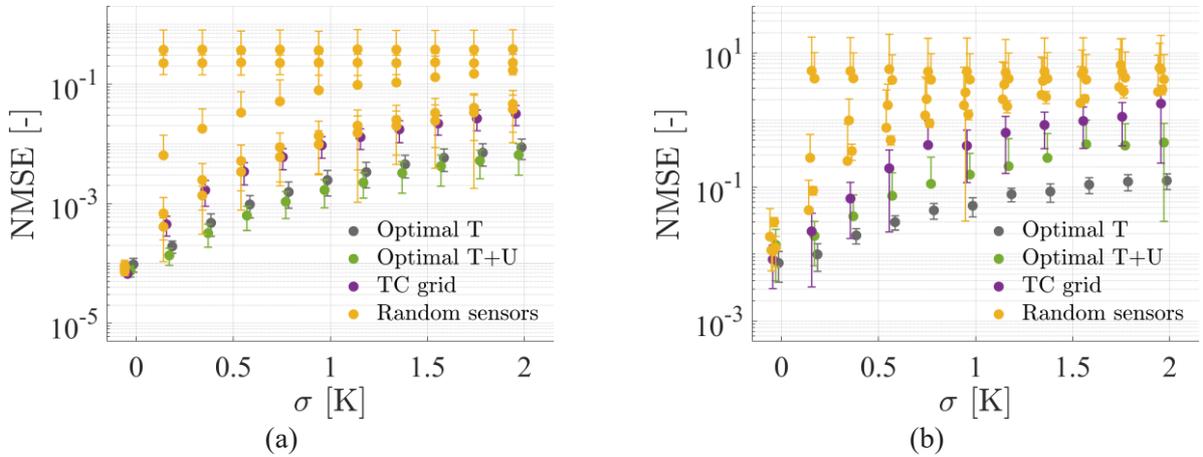

Figure 15. Reconstruction of (a) temperature and (b) velocity by sparse temperature measurements polluted by noise. Cases compared with different sensor arrangements. Error bar represents the standard deviation (std) of 10 repeated trainings.

Generally, consideration of noise in the input space is necessary for the practical application. To enhance the stability of the framework, regularization is required. Selection of the approximation algorithm and placement of sensors require a comprehensive study, and they are highly dependent on the specific case. Specifically, the use of linear or non-linear regression is determined by the underlying relationship between target values and inputs. This relationship is strongly correlated with the PDEs being solved as well as the covered boundary conditions when creating the training dataset.

Optimal sensor placement provides optimized sensor locations to maximize the measured information with a limited number of sensors. From the practical perspective, if a sensor is too close to near ones, it becomes impractical to implement. To achieve a similar prediction capability, extra sensors can be applied (e.g., TC grid). However, it should be noted that these additional sensors might also introduce noise and therefore worsen the robustness of the framework.

# 4 CASE2: CONDENSING JETS BY STEAM INJECTION THROUGH SPARGER

## 4.1 Description of the test case

Direct Contact Condensation (DCC) is applied in Light Water Reactors (LWRs) to prevent overpressure of the primary system. In Boiling Water Reactors (BWRs), steam in the reactor pressure vessel can be released into a large water pool, known as Pressure Suppression Pool (PSP), through multi-hole spargers of Automatic Depressurization System (ADS) in normal operation and accident scenarios. The PSP can be thermally stratified if the momentum created by condensed steam is insufficient to overcome its buoyancy. Pool stratification is regarded as a safety concern since it reduces the volume for heat storage and therefore results in a faster increase of pool surface temperature as well as the containment pressure compared to a completely mixed pool condition.

Validated codes with sufficient predictive capability to simulate realistic accident scenarios and resolve the interplay between phenomena, safety systems, and operational procedures are important for plant licensing, operation, and monitoring. To enable the modeling of thermal stratification and mixing phenomena in prototypic PSP conditions (i.e. long-term transient with many steam injection flow paths), the so-called "Effective Heat Source (EHS)" and "Effective Momentum Source (EMS)" models have been proposed [27]. Instead of explicitly resolving the interface between steam and liquid, the EHS/EMS models apply single-phase liquid with the same amount of momentum and energy sources to effectively reproduce the integral effect of DCC downstream [6].

The second case is condensing jets induced by steam injection through a multi-hole sparger in a water pool. Measurements were obtained from an integral effect test (SPA-T3) performed in PPOOLEX facility as shown in Figure 16. Experiment details and its post-test analysis are reported in [4][5]. During the test, steam was generated from a 1 MW steam generator and injected into the pool through a sparger. The sparger contained 32 holes arranged in 4 rings and each hole has an inner diameter of 8 mm. The pool was filled with 3m water at room temperature. The test consisted of two phases for the development of thermal stratification with relatively low steam flow rates and two phases for pool mixing with high flow rates. Four vertical lines of TCs (L1~L4) were positioned at different radial distances from the sparger pipe to monitor the global pool behavior. A TC grid (Figure 16b) arranged with 6 × 7 TCs was placed in front of a column of injection holes to record the local phenomena near the sparger. Readings obtained from selected TCs are presented in Figure 17. The sampling frequency for slow TCs labeled 1, 8, 29, 36~42 is 0.67Hz and for remaining fast TCs is 20Hz. The oscillated signal as shown in Figure 17b is primarily because of the rapid motion of fluid induced by the condensing steam.

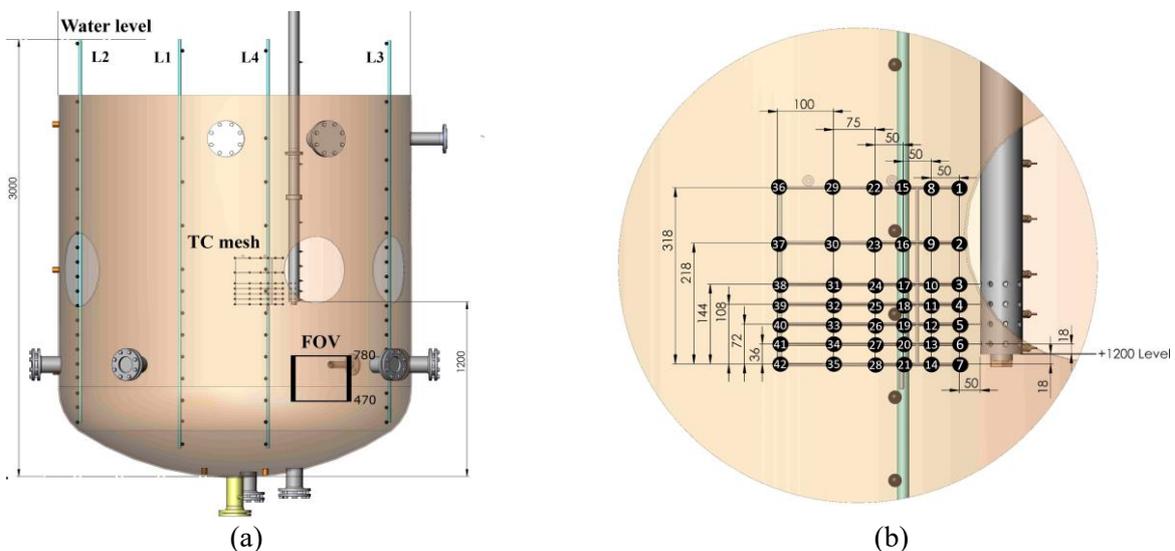

(a) (b)
Figure 16. (a) Overview and (b) TC grid of PPOOLEX facility for sparger tests.

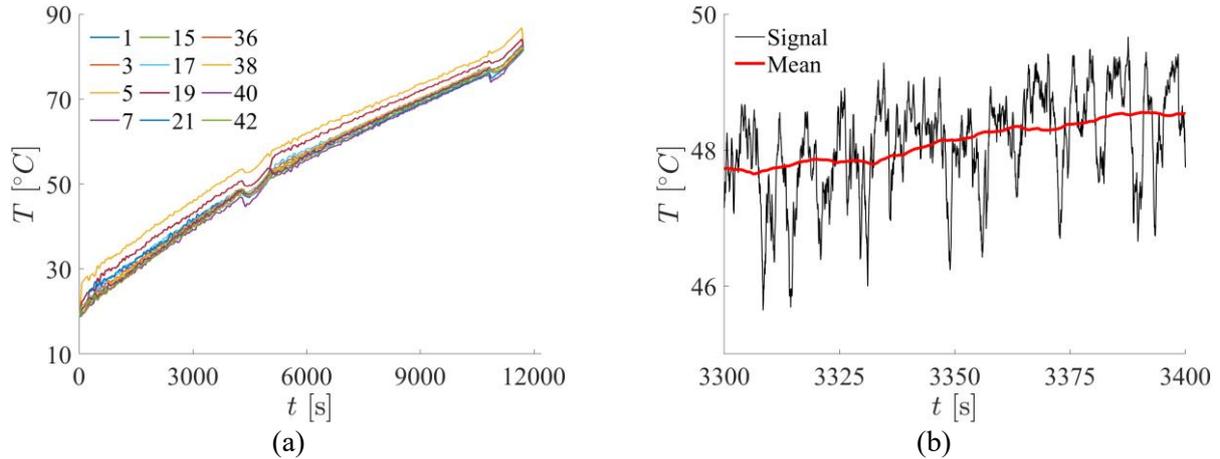

Figure 17. Evolution of (a) temperature recorded from selected thermocouples of TC grid and (b) raw and averaged signals of TC #5 in PPOOLEX SPA-T3. Temperature was time-averaged over 50s.

The training library was created by collecting simulation results from CFD validation against this experiment. Generally, the numerical scheme applied an incompressible, transient, single-phase solver as described by Eqs. (1)~(3). The effect of buoyancy was considered by using temperature-dependent water properties. Turbulence was solved by using the RANS $k - \omega$ BSL model and the energy equation was turned on. Implementation of EHS/EMS models was achieved by imposing boundary conditions on injection holes (Figure 18b) with the same amount of effective momentum and heat as injecting steam. Modeling details are reported in [28].

PIV measurement of steam injection through a sparger into a water pool indicates a high level of turbulence in the vicinity of the sparger [3][4]. This turbulence is believed to be caused by the rapid motion of the steam/water interface induced by condensing bubbles. Good agreements on global pool behavior and local flow characteristics were achieved when a turbulence source was imposed at the region where condensation occurs [6]. This additional turbulence source serves to diffuse the jets so that pool stratification is more likely to develop. However, the selection of proper parameters defining this turbulent source is currently the major source of uncertainty in pool modeling.

In the work done by authors [28], a parametric study with ~150 simulations was conducted to calibrate and validate the proposed pool modeling approach against PPOOLEX SPA-T3 test. The major variation among these cases is the parameters used to distribute the turbulence source induced by steam condensation. These parameters determine the (i) magnitude, (ii) imposing region, (iii) azimuthal, and (iv) radial distribution profiles of the turbulence source. Each case wrote the velocity and temperature profiles on the same slice plane where the TC grid was placed (Figure 16) with a frequency of 100s, yielding 10~120 snapshots depending on its running time. Consequently, the dataset consists of 6800 snapshots of complete velocity and temperature profiles.

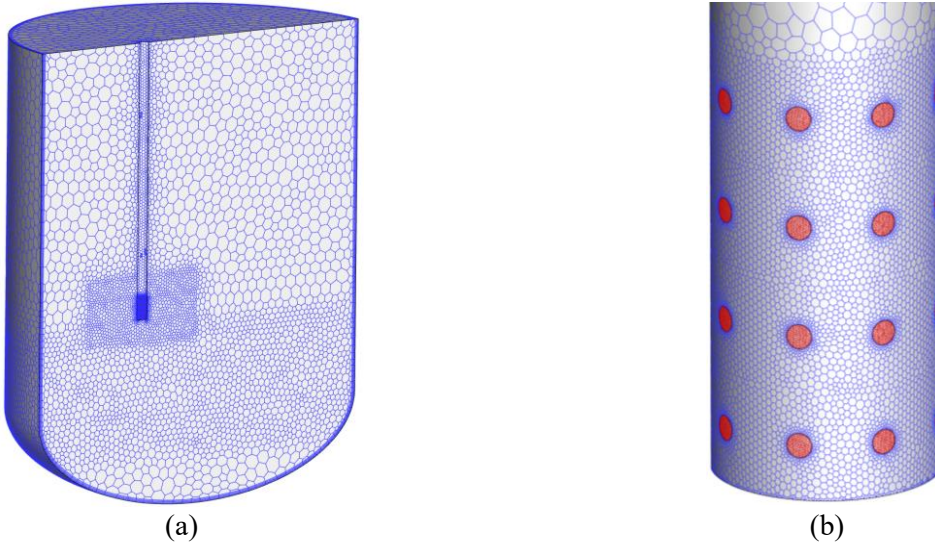
Figure 18. Computation domain using polyhedral cells of (a) pool overview and (b) sparger details.

## 4.2 Results and discussions

The full dimension of U and T fields can be optimally described by 10 and 74 modes, respectively, which account for 99.99% and 99.9% of their variability. The velocity manifold shows greater complexity than the temperature field. It contains several dominant modes followed by a large number of relatively small but comparable significance modes. Additionally, the primary and secondary modes of temperature and velocity exhibit similar features, either convection- or diffusion-dominant, compared to the turbulent planar jet as presented in Section 3.2.1.

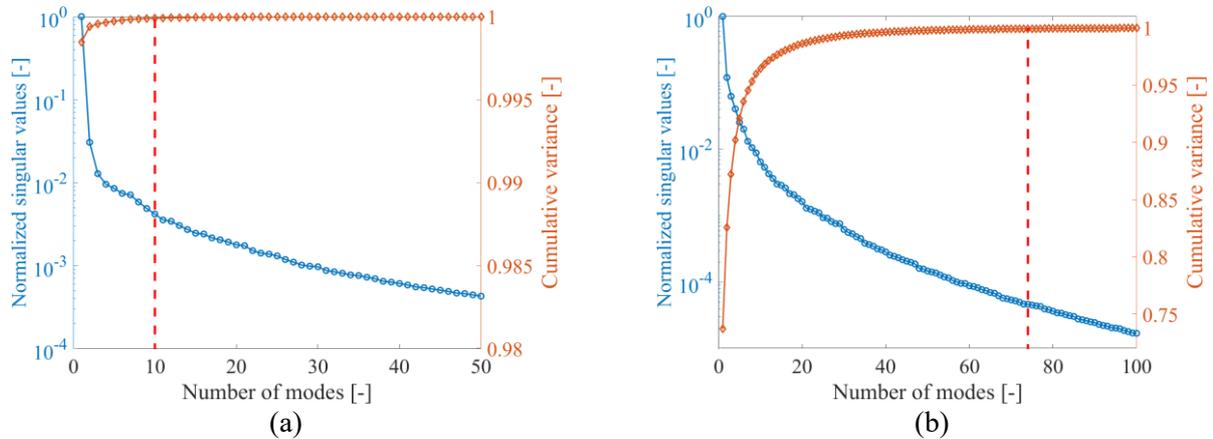
Figure 19. Cumulative variance and normalized singular values as a function of the number of modes for (a) temperature and (b) U velocity fields of steam injection case.

According to the discussion presented in Section 3.2.4, the reconstruction of both temperature and velocity fields requires not only sparse temperature data but also velocity (or its equivalent, such as mass flow rate) at the orifice. In this case, the temperatures from the positions as arranged in Figure 16b together with the effective velocity estimated by the EMS model [6][28] were extracted as inputs. This effective velocity also served as the imposed boundary in the CFD simulations.

To enable a robust approximation approach to map these inputs to latent coefficients of their corresponding modes (Figure 19), we conducted a comprehensive study, focusing on the impacts of (i) linear regression/neural network and (ii) lasso/ridge regularizations on the performance. Eventually, the neural network with 3 hidden layers, each containing 10 nodes and a ridge regularization term with $\gamma = 0.2$ [23] was selected due to its overall good reconstruction performance against data with small noise (i.e. $\sigma \leq 0.4$). The coefficients of primary and secondary modes of the velocity field predicted by this neural network with noise-free input are compared with its reference as shown in Figure 20. Figure 21 presents example snapshots of the reconstructed temperature and velocity fields. The relatively small deviation shows a good predictive capability of the proposed framework.

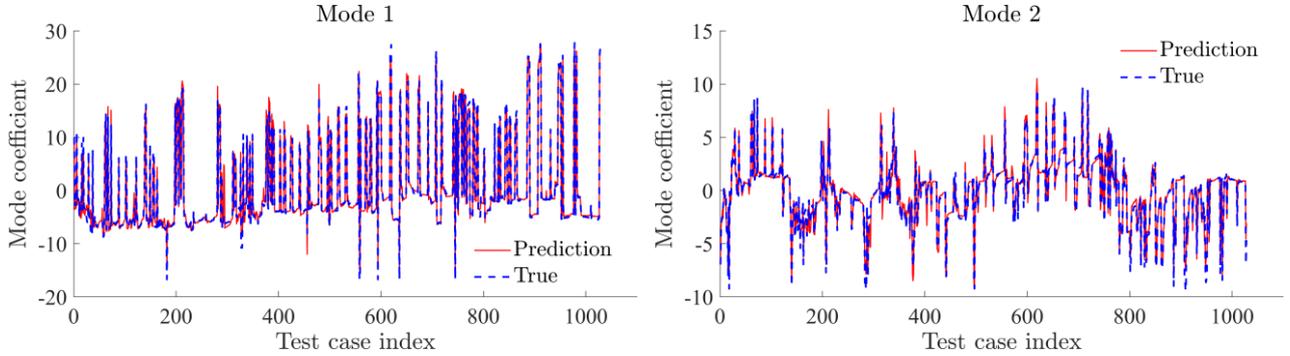

Figure 20. Comparison of coefficients of velocity latent space decoded by POD (True) and predicted by the neural network with $\gamma = 0.2$.

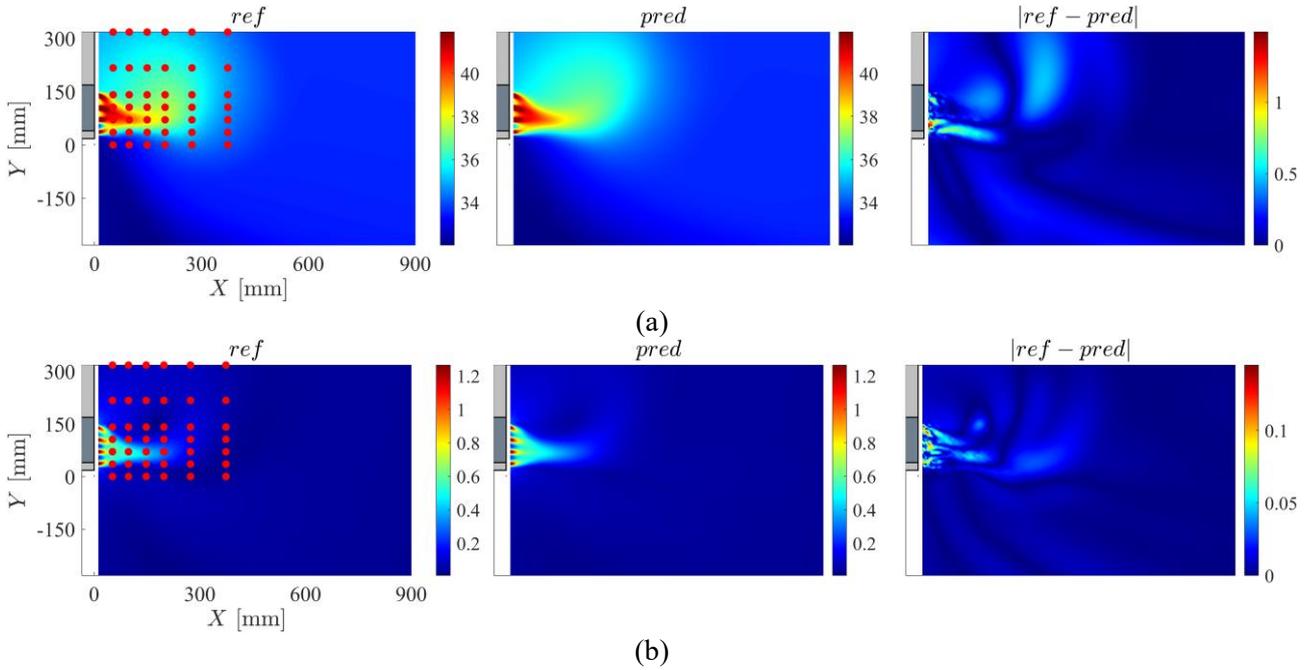

(a)

(b)

Figure 21. Reconstructed of (a) temperature and (b) velocity magnitude on the testing dataset with low steam flow flux by sparse temperatures obtained in CFD and effective velocity.

Reconstructions using sparse temperatures measured in PPOOLEX SPA-T3 and effective velocity estimated by injection conditions are displayed in Figure 22 wherein a complete video with all snapshots can be found in [Appendix](). Given that the training data is collected from CFD simulations using the RANS turbulence solver in combination with the EHS/EMS models, TCs readings were time-averaged over 50s (Figure 17b) to omit the small-scale effects such as flow fluctuation induced by the motion of the steam/water interface.

Although it is impractical to evaluate the prediction performance due to the unavailability of measurements of both fields in full dimension, we can still observe that the two distinct flow patterns, i.e., buoyancy-driven and inertial-dominant flows, are well captured in temperature reconstruction and partly represented through velocity reconstruction. An unremarkable separation is observed downstream of the velocity field which is not physical in practical scenarios. This deviation suggests the necessity for involving supplementary information for velocity reconstruction. For instance, this information could be the downstream velocity measured by PIV located in a way such that the impact of steam/water interaction could be mitigated.

Given that the full temperature measurement is not available for all PPOOLEX experiments, several TC readings are hidden as the testing data. The training and testing were conducted by applying values from the remaining sensors as the input. Temperatures measured by these hidden TCs are compared with values extracted from the reconstructed temperature field as shown in Figure 23. Generally, a good agreement is observed for most of the TCs except TC #12 in which the prediction is slightly larger than the measurement.

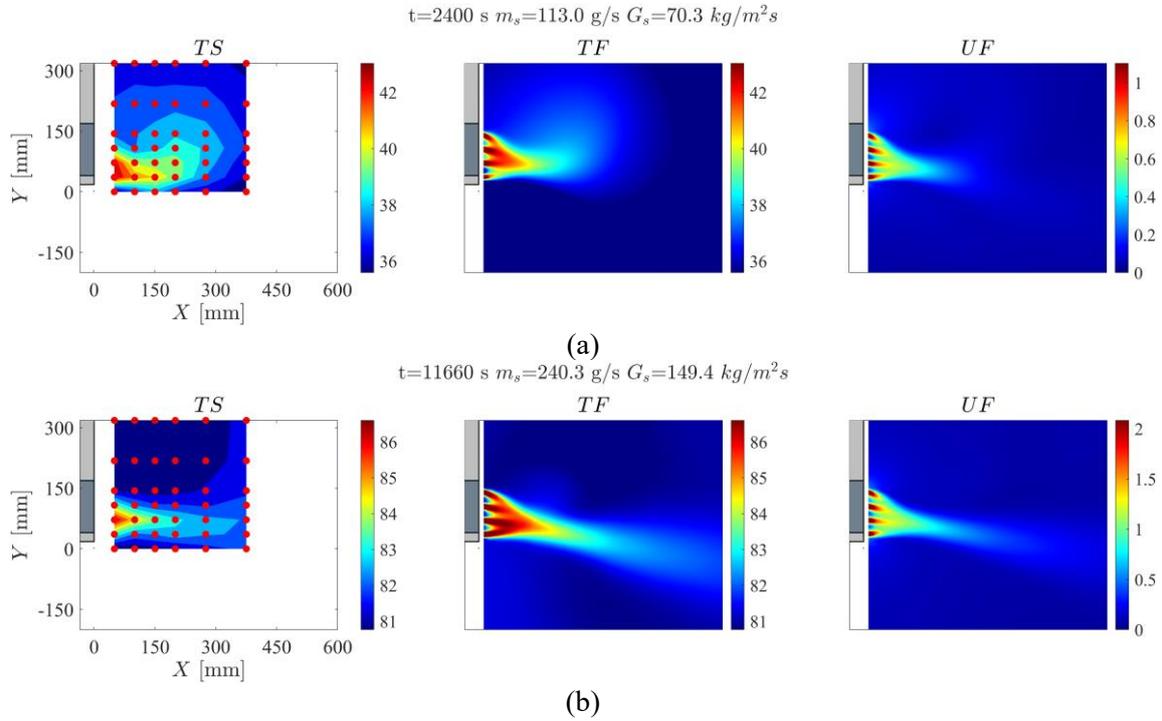

Figure 22. Reconstructed temperature (TF) and velocity magnitude (UF) by sparse temperatures measured in the test and effective velocity in (a) low and (b) high steam flux.

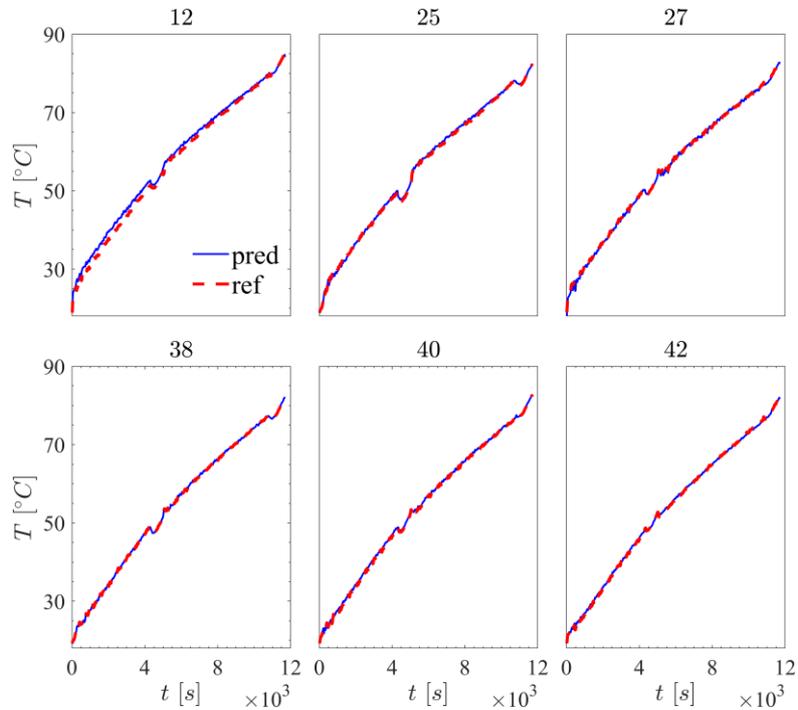

Figure 23. Comparison of hidden TCs between values measured in PPOOLEX SPA-T3 (ref) and predictions (pred) by the proposed data-driven framework.

# 5       Conclusions and outlooks

Direct measurement of velocity field by means of optical techniques such as PIV is challenging in thermal-hydraulic experiments with multi-phase interaction or non-transparent fluid. Phenomena interpretation as well as code development and validation typically rely on the measurements obtained through sparsely distributed probes, e.g., thermocouples, and pressure transducers. In this paper, we proposed a data-driven modeling approach to reconstruct turbulent flow from temperature measured by sparsely distributed TCs. The framework consists of (i) an encoder to encode the full-dimension field from its latent space and (ii) a mapping function to estimate the latent space coefficients through sparse measurements. This approach has been tested on a single-phase planar jet and steam condensing jets issued from a multi-hole sparger. The main findings and conclusions are summarized below:

- The POD, compared to the autoencoder, yields a better performance for mode reduction in which the high-dimensional temperature and velocity datasets of these two cases can be decomposed into 10~74 modes depending on the data complexity and required variability.
- Fully connected neural network and linear regression were implemented to approximate the latent space coefficients through sparse sensor data. Reconstruction of temperature can be achieved by using only a few temperature sensors while velocity reconstruction requires additional information such as velocity at the orifice or downstream.
- Both approximation approaches provide a good predictive capability against noise-free input. To enable a robust prediction towards input with noise, regularization is necessary. The FNN with ridge regularization term exhibits a relatively better performance in less noisy input while the linear regression with ridge regularization term outperforms in case of large noise.
- Sensors arranged by optimal sensor placement have greater robustness against noise input compared to randomly sampled sensors. However, these optimized sensors might be impractical to implement, and adjustment is required. Increasing the number of sensors, although providing more information, can also introduce additional noise, reducing the robustness of the model.
- Application of the proposed framework to a steam injection test shows promising results. The accuracy of reconstructed temperature profiles is accessed by the readings from hidden sensors. The buoyancy-driven and internal dominant flows are well predicted.

Due to the lack of velocity measurement in PPOOLEX tests, it becomes impossible to validate the velocity reconstruction. Experiments conducted in PANDA facility provided PIV recordings but without temperature measured in the vicinity of the injection holes. Further work needs to be done to simultaneously measure the velocity and temperature and it is under discussion in the following PANDA project. It would be interesting to implement the optimal sensor placement technique for the pre-test analysis. The simulation results obtained from scoping analysis with varied boundary conditions and parameterized PDEs can be used as the training dataset. The designed sensors are expected to provide measurements that can mostly reduce the uncertainties of the modeling parameters. Moreover, a further study could assess the feasibility of the application of this framework for thermal-hydraulic experiments using liquid metals such as TALL-3D [1].

## ACKNOWLEDGEMENTS

The authors are thankful for the feedback and discussions with the colleagues at the PPOOLEX facility in LUT and support from the NKS (Nordic Nuclear Safety Research) THEOS project, and Finnish SAFIR project. The authors would also like to thank Prof. Ricardo Vinuesa Motilva and Mr. Yuning Wang from KTH for the discussion of data-driven modeling and the potential application of PINN.

## DATA AVAILABILITY

The code and dataset are available on GitHub at: https://github.com/xichenggege/SparseT2V.git. The appendix is available at SparseT2V_paper_share.